# Bayesian fusion and multimodal DCM for EEG and fMRI

**Huilin Wei[1,2], Amirhossein Jafarian[2], Peter Zeidman[2], Vladimir Litvak[2], Adeel Razi[2,3,4,5], Dewen Hu[1]\*, Karl J. Friston[2]\***

[1]College of Intelligence Science and Technology, National University of Defense Technology, Changsha, Hunan, People's Republic of China
[2]Wellcome Centre for Human Neuroimaging, UCL Institute of Neurology, University College London, London, United Kingdom
[3]Turner Institute for Brain and Mental Health, School of Psychological Sciences, Monash University, Clayton, Australia
[4]Monash Biomedical Imaging, Monash University, Clayton, Australia
[5]Department of Electronic Engineering, NED University of Engineering and Technology, Karachi, Pakistan

**\* Correspondence to:**
Prof. Karl J. Friston, Wellcome Centre for Human Neuroimaging, UCL Institute of Neurology, University College London, 12 Queen Square, London, WC1N 3BG, United Kingdom. E-mail address: k.friston@ucl.ac.uk (K.J. Friston).

Prof. Dewen Hu, College of Intelligence Science and Technology, National University of Defense Technology, 109 Deya Road, Changsha, Hunan, 410073, People's Republic of China. E-mail address: dwhu@nudt.edu.cn (D.W. Hu).

## Abstract

This paper asks whether integrating multimodal EEG and fMRI data offers a better characterisation of functional brain architectures than either modality alone. This evaluation rests upon a dynamic causal model that generates both EEG and fMRI data from the same neuronal dynamics. We introduce the use of *Bayesian fusion* to provide informative (empirical) neuronal priors – derived from dynamic causal modelling (DCM) of EEG data – for subsequent DCM of fMRI data. To illustrate this procedure, we generated synthetic EEG and fMRI timeseries for a mismatch negativity (or auditory oddball) paradigm, using biologically plausible model parameters (i.e., posterior expectations from a DCM of empirical, open access, EEG data). Using model inversion, we found that Bayesian fusion provided a substantial improvement in marginal likelihood or model evidence, indicating a more efficient estimation of model parameters, in relation to inverting fMRI data alone. We quantified the benefits of multimodal fusion with the information gain pertaining to neuronal and haemodynamic parameters – as measured by the Kullback-Leibler divergence between their prior and posterior densities. Remarkably, this analysis suggested that EEG data can improve estimates of haemodynamic parameters; thereby furnishing proof-of-principle that Bayesian fusion of EEG and fMRI is necessary to resolve conditional dependencies between neuronal and haemodynamic estimators. These results suggest that Bayesian fusion may offer a useful approach that exploits the complementary temporal (EEG) and spatial (fMRI) precision of different data modalities. We envisage the procedure could be applied to any multimodal dataset that can be explained by a DCM with a common neuronal parameterisation.

**Keywords:** *Multimodal data, Bayesian fusion, dynamic causal modelling, canonical microcircuit neural mass model, laminar architecture, information gain*

## Introduction

Dynamic causal modelling (DCM) was initially proposed to infer effective connectivity from fMRI data using parsimonious one-state-per-region model generators of neuronal activity (Friston et al., 2003; Li et al., 2011; Friston et al., 2014). DCM was subsequently developed to characterise the canonical microcircuits – with laminar structure – underlying event related or induced responses measured with EEG or MEG (David et al., 2006; Chen et al., 2008; Friston et al., 2012). In general, dynamic causal models have a neuronal part and a measurement (or observation) part; for example, the haemodynamic model in DCM for fMRI. These models specify how experimental stimuli induce neuronal dynamics and then how neuronal responses cause observable responses, such as EEG sensor data and BOLD signals. A recent development, the canonical microcircuit DCM for fMRI, offers a canonical microcircuit neural mass model that underwrites neuronal dynamics (Friston et al., 2017). This neuronal model is based on successive model developments and comparison using DCM for M/EEG (Pinotsis et al., 2012; Moran et al., 2013; Pinotsis et al., 2013) and has the following aspects: first, it provides a generative or forward model of measurable signals that retains a degree of neurobiological plausibility. Experimental inputs (e.g., visual stimuli, attention modulation, etc.) perturb neuronal dynamics within coupled canonical microcircuits, each comprising a laminar specification of cell types and their interconnectivity (Thomson and Bannister, 2003). These neuronal populations elaborate forward and backward connections to model cortical hierarchies (Bastos et al., 2012; Bastos et al., 2015). Second, it equips a haemodynamic model (Friston et al., 2000) with a laminar specific neural drive. These neuronal afferents further characterise a laminar specification of neurovascular architecture that may be useful for modelling high resolution fMRI (Heinzle et al., 2016; Markuerkiaga et al., 2016; Scheeringa et al., 2016; Huber et al., 2018; Kashyap et al., 2018; Uludag and Blinder, 2018).

However, from a computational perspective, inverting such neural mass models using fMRI data is inefficient, due to the poor temporal resolution of this imaging modality. In contrast, there is more information in electromagnetic data about underlying neuronal and synaptic parameters, which means that more parameters can be estimated efficiently; i.e., with greater conditional certainty (David et al., 2006). A straightforward multimodal inversion strategy is to treat the neuronal and haemodynamic model parameters as conditionally independent, i.e., they can be efficiently estimated separately using M/EEG and fMRI data (Rosa et al., 2011). Along these lines, a recently introduced method within the DCM framework begins by deriving neuronal responses from a neural mass model of M/EEG data, which are then used to drive a haemodynamic model fitted to fMRI data (Jafarian et al., 2019). This approach provides an efficient pipeline for multi-modal integration. However, our priority here was to estimate the information gain on neuronal and haemodynamic parameters afforded by EEG and/or fMRI data, which precluded the use of independence assumptions about neuronal and haemodynamic parameters. Therefore, we instead performed multimodal *Bayesian fusion*, in which the (posterior) estimates of the parameters from one imaging modality were used as the priors for the other. This enables all parameters to be informed by both the M/EEG and fMRI data. Specifically, under this scheme, one can first employ electrophysiological measurements such as M/EEG to constrain posterior estimates of canonical microcircuitry, then take these posteriors as (empirical) neuronal priors for a subsequent inversion using fMRI data. This second inversion furnishes (electrophysiologically) informed estimates of regionally specific haemodynamic parameters. This form of fusion, which rests upon a DCM with a common neuronal parameterisation, is known as *Bayesian fusion via Bayesian belief updating* (Friston et al., 2017). The resulting multimodal Bayesian fusion thereby provides a characterisation of functional brain architectures that properly combines two sources of information that are well-resolved either in time (electromagnetic) or space (fMRI).

In this technical note, we assess the benefits of multimodal Bayesian fusion when making inferences about neuronal and haemodynamic responses using a canonical microcircuit DCM. The following paper comprises three sections. The first section introduces the theoretical foundations of multimodal DCM and Bayesian fusion. In this section, we combine the two strands of Bayesian modelling to provide a generative model of DCM suitable for EEG-fMRI fusion. We describe the generative model of M/EEG and fMRI data in detail, and introduce a pipeline for implementing multimodal Bayesian fusion. The second section provides an illustrative application of Bayesian fusion using a synthetic multimodal dataset. We first call on the generative model described in the previous section and biologically plausible parameters (i.e., posterior expectations based upon a standard DCM analysis of empirical EEG data) (Garrido et al., 2009a) to generate EEG and fMRI timeseries for a mismatch negativity paradigm. These simulations illustrate the fact that both electromagnetic and haemodynamic responses are observable consequences of hidden neuronal activity. Having simulated multimodal data (with appropriate levels of measurement noise), we implement Bayesian fusion; specifically, we invert the model with EEG data and use the ensuing posterior densities as priors (on the shared neuronal parameters) for an inversion of the fMRI data. We also conduct model inversion using just fMRI data, for later comparison with Bayesian fusion. The quality of the ensuing inversions is assessed in terms of the log evidence (as approximated by the Variational free energy). We hypothesised that the unimodal fMRI analyses would be less able to identify detailed changes in microcircuitry (i.e., mismatch effects), whereas the Bayesian fusion would provide more precise estimates – an accompanying increase in the efficiency of parameter estimation. The last section assesses the usefulness of EEG and fMRI data in terms of their relative information; i.e., information gain as measured by the Kullback-Leibler divergence (Kullback and Leibler, 1951; Zeidman et al., 2018), pertaining to neuronal and haemodynamic parameters associated with the two data modalities and their fusion. This *Bayesian data comparison* provides a quantitative way to evaluate the benefits of multimodal Bayesian fusion – and the degree to which distinct data modalities resolve uncertainty about unknown model parameters.

## Methods and materials

### Dynamic causal modelling for EEG-fMRI fusion

Dynamic causal modelling suitable for multimodal fusion adopts a common neuronal model but modality specific observation models, to explain multiple measurements (e.g. M/EEG, fMRI timeseries) as the observable consequence of neuronal activity. Fig. 1 provides a schematic summary of the generative model of DCM for EEG-fMRI fusion, in which the neuronal part is described by the canonical microcircuit neural mass model (Friston et al., 2017), while the observation part is a standard electromagnetic forward model for EEG (Kiebel et al., 2006) and the haemodynamic model for fMRI (Stephan et al., 2007), respectively. All corresponding variables are listed in Table 1 and Table 2.

The right panel of Fig. 1 illustrates the structure of the canonical microcircuit and the equations modelling neuronal dynamics. The canonical microcircuit (per region or source) comprises four neuronal populations in distinct cortical layers, corresponding to spiny stellate cells in the granular layer (denoted by population 1), superficial pyramidal cells in the supragranular layer (population 2), deep pyramidal cells in the infragranular layer (population 4), as well as inhibitory interneurons that constitute the only inhibitory population (population 3). The four populations are coupled via inter- and intralaminar intrinsic connectivity, where light blue connections in the figure denote intrinsic excitatory connectivity (mediated by intrinsic connectivity strength $a_{ij}$, $a_{ij} > 0$, $ij$ indicates the direction from $j$ to $i$). The pink connections represent intrinsic inhibitory connectivity (mediated by $a_{ij}$, $a_{ij} < 0$), and the pink circles denote the recurrent (self) connections that are universally inhibitory (mediated by self-inhibition strength $G_i$). In addition to the intrinsic (within region) connectivity, there exists extrinsic (between region) connectivity with empirically established laminar-specific origins and targets (dark blue connections in Fig. 1). These extrinsic connections can be divided into forward and backward connections in cortical hierarchies, where extrinsic forward connections arise from superficial pyramidal cells of a lower-level region and target spiny stellate cells and deep pyramidal cells (mediated by extrinsic connectivity strength $A_{kl}^{(12)}, A_{kl}^{(42)}$, as shown in Fig. 1), while extrinsic backward connections arise from deep pyramidal cells of a higher-level region and target superficial pyramidal cells and inhibitory interneurons (mediated by $A_{kl}^{(24)}, A_{kl}^{(34)}$). Experimental inputs can directly drive or modulate neuronal responses. The experimental driving effects are determined by the parameters of a $C$ matrix, which target spiny stellate cells in each region. The experimental modulatory effects are specified by changes in intrinsic and/or extrinsic connectivity strength, which are parameterised as $B$ matrices.

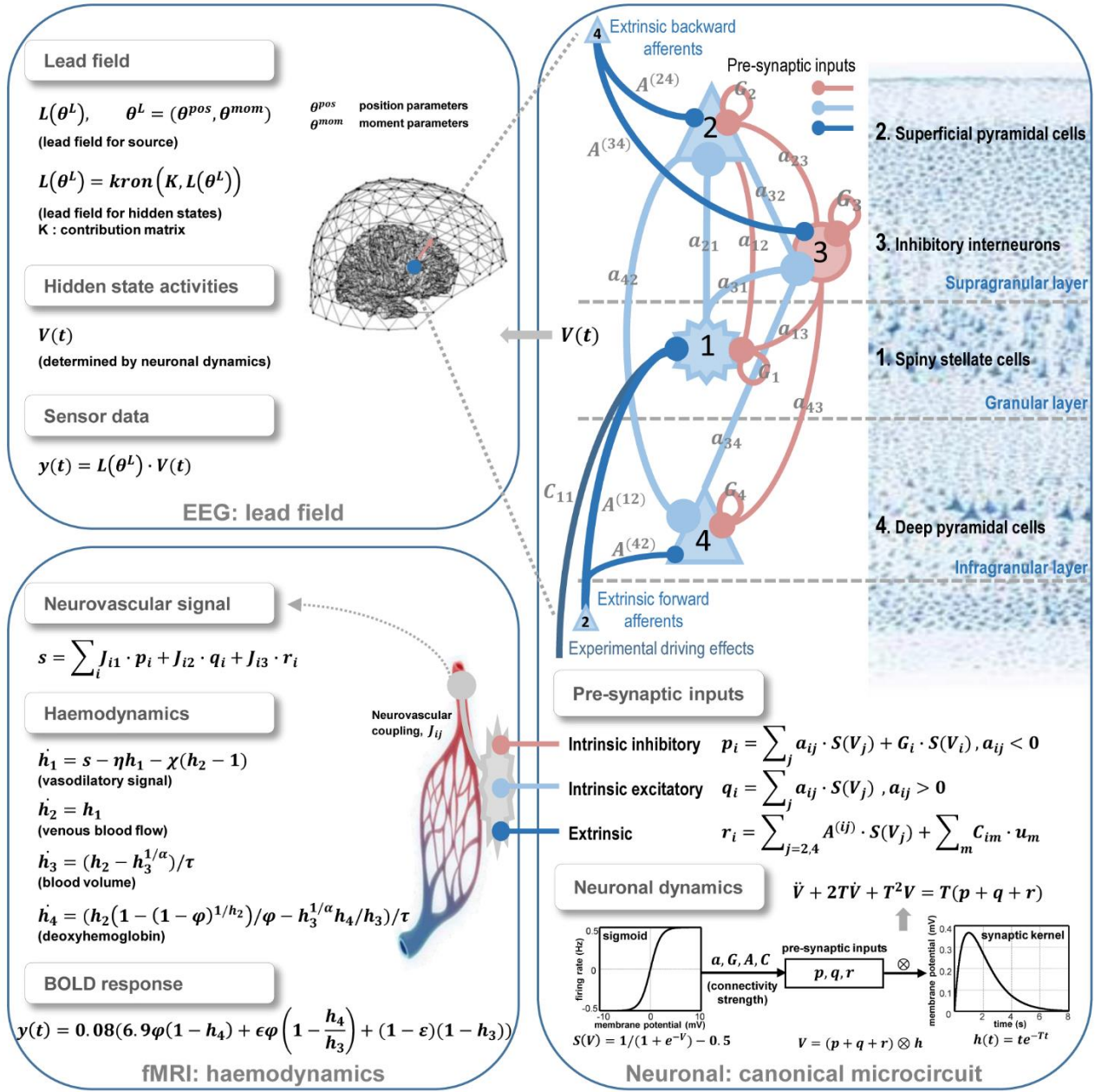

**Fig. 1.** The generative model of DCM for EEG-fMRI fusion (per region). This model comprises two parts: the common neuronal part (right panel), and a modality-specific observation part (left panel). Responses to experimental inputs – perturbing neuronal dynamics – are modelled with a canonical microcircuit. This microcircuit comprises four neuronal populations in distinct cortical layers, each population receives three sorts of pre-synaptic inputs (scaled by respective connectivity strength parameters), corresponding to intrinsic inhibitory afferents (denoted by pink lines and circles), intrinsic excitatory afferents (light blue lines), and extrinsic (forward and backward) afferents (dark blue lines). Each population is equipped with a hidden state, whose dynamics are described by the second-order ordinary differential equation in the figure. The observation models are a standard lead field for EEG and the haemodynamic model for fMRI. For EEG, the lead field (determined by some EEG spatial parameters) is a spatial linear mapping that transforms neuronal activity into the EEG sensor data. For fMRI, the observation model comprises neurovascular coupling and haemodynamics. The neurovascular coupling describes how neuronal dynamics induce a vasoactive signal. The neurovascular signal is generated via a linear combination of the pre-synaptic inputs (scaled by the neurovascular coupling parameters). The subsequent haemodynamics describe how neurovascular signal drives a standard haemodynamic model to generate the final BOLD response. Please see Table 1 and Table 2 for a description and parameterisation of the variables.

Based on the architecture of the canonical microcircuit, the neuronal dynamics are described by second-order ordinary differential equations of motion for $V(t)$ as shown in Fig. 1, where $V(t)$ are the mean transmembrane potentials of populations (i.e., hidden neuronal states). In more detail, experimental inputs change hidden neuronal states, where a sigmoid activation function transforms the transmembrane potentials of populations into firing rates $S(V)$, which are then weighted by intrinsic $(a, G)$ and extrinsic $(A, C)$ connectivity parameters to form the pre-synaptic inputs to each population. According to the categories of intrinsic and extrinsic connections, the pre-synaptic inputs at each population can be divided into three sorts, corresponding to intrinsic inhibitory afferents (denoted by $p$, pink colour), intrinsic excitatory afferents ($q$, light blue), and extrinsic afferents ($r$, dark blue). The subsequent dynamics (Jansen and Rit, 1995; Moran et al., 2013) effectively convolve pre-synaptic inputs with a synaptic kernel to depolarise post-synaptic populations, where $T$ denotes population-specific post-synaptic rate constants. Conduction delays (denoted by $D$) are parameterised with values that correspond to the time taken for axonal propagation between regions (~16 ms) (Friston et al., 2017). All the above parameters, which we collectively refer to the *neuronal* parameters, are free parameters of the model that can be optimised during model inversion for both EEG and fMRI data (see Table 1).

The left panel of Fig. 1 shows the modality specific observation models corresponding to EEG and fMRI data. For EEG, the observation model is a spatial linear mapping that transforms the estimated source (neuronal) activities $V(t)$ into the EEG sensor (scalp) data $y(t)$. This mapping or matrix operator is called the lead field $L(\theta^L)$ – and is determined by some spatial parameters including three location and three orientation (or moment) parameters $\theta^L = (\theta^{pos}, \theta^{mom})$ (Kiebel et al., 2006; Kiebel et al., 2007a). A contribution matrix $K$ is used to weight the depolarisation of all populations to the measurable (dendritic) signal from each source. The dominant contribution is usually from the pyramidal population, which is the prominent source that can be detected remotely on the scalp surface in M/EEG (because the spatial arrangement of pyramidal cell dendrites is perpendicular to the cortical surface) (David et al., 2006; Kiebel et al., 2006).

For fMRI, the observation model comprises neurovascular coupling and haemodynamics. Neurovascular coupling describes how neuronal dynamics induce a vasoactive signal; here, we assume that each sort of pre-synaptic input to a neuronal population (i.e., intrinsic inhibitory, intrinsic excitatory and extrinsic) is accompanied by a collateral input of the same strength to nearby astrocytes. The neurovascular signal (denoted by $s$ in Fig. 1) is then generated via a linear combination of all pre-synaptic drives in a region (scaled by neurovascular coupling parameters $J_{ij}$) (Carmignoto and Gomez-Gonzalo, 2010; Friston et al., 2017). The subsequent haemodynamics describe how neurovascular signal drives a standard haemodynamic model (Friston et al., 2000; Stephan et al., 2007) to generate the BOLD response.

In brief, a neurovascular signal $s$ enters the haemodynamic model and drives a vasodilatory signal $h_1$, blood flow $h_2$, which responds to the vasodilatory signal and causes changes in blood volume $h_3$ and deoxyhemoglobin $h_4$. Finally, the observed BOLD signal $y(t)$ is generated by a nonlinear function of volume and deoxyhemoglobin (Buxton et al., 2004). The parameters in the haemodynamic state equations and neurovascular coupling parameters – which we collectively refer to the *haemodynamic* parameters – are described in detail in Table 2. These modality-specific parameters of the observation models can only be estimated with the corresponding data modality.

**Table 1 Neuronal parameters**

| | Description | Parameterisation | Prior |
|---|---|---|---|
| $T_i^{(k)}$ | Synaptic rate constant of the $i$-th neuronal population in region $k$ | $\exp(\theta_T) \cdot T_i$ <br> $T = [256, 128, 16, 32]$ | $p(\theta_T) = N(0, \frac{1}{16})$ |
| $a_{ij}^{(k)}$ | Intrinsic connectivity from population $j$ to population $i$ in region $k$ | $\exp(\theta_S) \cdot a$ <br> $a = \begin{bmatrix} -8 & -4 & -4 & 0 \\ 4 & -8 & -2 & 0 \\ 4 & 2 & -4 & 2 \\ 0 & 1 & -2 & -4 \end{bmatrix}$ <br> where $S$ is the intrinsic gain | $p(\theta_S) = N(0, \frac{1}{16})$ |
| $G_i^{(k)}$ | Self-inhibition of the $i$-th neuronal population in region $k$ | $\exp(\theta_G) \cdot a_{ii}$ | $p(\theta_G) = N(0, \frac{1}{16})$ |
| $A_{kl}^{(ij)}$ | Extrinsic connectivity from population $j$ in region $l$ to population $i$ in region $k$ | $\exp(\theta_A) \cdot A^{(12)}, A^{(12)} = 1024$ <br> $\exp(\theta_A) \cdot A^{(42)}, A^{(42)} = 512$ <br> $\exp(\theta_A) \cdot A^{(24)}, A^{(24)} = 512$ <br> $\exp(\theta_A) \cdot A^{(34)}, A^{(34)} = 512$ | $p(\theta_A) = N(0, \frac{1}{8})$ |
| $B_{klm}$ | Change in connectivity caused by the $m$-th input | $\theta_B$ | $p(\theta_B) = N(0, \frac{1}{8})$ |
| $C_{im}^{(k)}$ | Direct driving effect of the $m$-th input on population $i$ in region $k$ | $\exp(\theta_C)$ | $p(\theta_C) = N(0, \frac{1}{32})$ |
| $D_{kl}$ | Conduction delay from region $l$ to region $k$ | $\exp(\theta_D)$ | $p(\theta_D) = N(0, \frac{1}{32})$ |

**Table 2 Haemodynamic parameters**

| | Description | Parameterisation | Prior |
|---|---|---|---|
| $J_{ij}$ | Neurovascular coupling of the $i$-th neuronal population ($j$=1,2,3) | $\theta_J$ | $p(\theta_J) = N(0, \frac{1}{16})$ |
| $\eta^{(k)}$ | Rate of vasodilatory signal decay per sec in region $k$ | $0.64 \cdot \exp(\theta_\eta)$ | $p(\theta_\eta) = N(0, \frac{1}{256})$ |
| $\chi$ | Rate of flow-dependent elimination | $0.32 \cdot \exp(\theta_\chi)$ | $p(\theta_\chi) = N(0, 0)$ |
| $\tau^{(k)}$ | Rate of haemodynamic transit per sec in region $k$ | $2.00 \cdot \exp(\theta_\tau)$ | $p(\theta_\tau) = N(0, \frac{1}{256})$ |
| $\alpha$ | Grubb's exponent | $0.32 \cdot \exp(\theta_\alpha)$ | $p(\theta_\alpha) = N(0, 0)$ |
| $\varepsilon$ | Intravascular: extravascular ratio | $1.00 \cdot \exp(\theta_\varepsilon)$ | $p(\theta_\varepsilon) = N(0, \frac{1}{256})$ |
| $\varphi$ | Resting oxygen extraction fraction | $0.40 \cdot \exp(\theta_\varphi)$ | $p(\theta_\varphi) = N(0, 0)$ |

*Multimodal Bayesian fusion*

Having a generative model for multimodal fusion means that one can use multiple data modalities to inform the parameters of the same model. Generally, the data modalities should have a complementary nature; namely, one modality would provide more precise constraints on the parameters that are estimated less efficiently using the other modality – and *vice versa*. Therefore, in EEG-fMRI fusion, we expect the high spatial resolution of fMRI to locate functionally specialized regions that constitute the DCM, while relying on the high temporal resolution of EEG to provide precise estimates of canonical microcircuitry in terms of neuronal connectivity parameters. Furthermore, the fMRI modality can constrain estimates of haemodynamic parameters. These observations license the following pipeline for fusing EEG and fMRI (i.e., Bayesian fusion) with canonical microcircuit DCM:

1. Identify regions significantly activated by an experimental paradigm, using whole brain (SPM) analysis of fMRI data. These regions constitute the network architecture for subsequent DCM analysis.
2. Conduct a DCM analysis of EEG data by setting the prior location of sources (i.e., $\theta^{pos}$ in Fig. 1) to the location of the activated regions from step 1, to obtain the posterior densities over neuronal parameters.
3. Conduct a subsequent DCM analysis of fMRI data by replacing the uninformative priors over neuronal parameters (Gaussian shrinkage priors; see Table 1) with the posterior means and covariances from step 2, to inform posterior estimates of haemodynamic parameters.

In the above procedure, we use both EEG and fMRI to provide complementary constrains on model parameters. In particular, we rely on EEG data to provide precise or informative posterior estimates over neuronal parameters, in step 3 – known formally as Bayesian belief updating – the estimation of haemodynamic parameters should benefit due to the resolution of conditional dependencies between neuronal and haemodynamic parameters. This increase in the efficiency of parameter estimation is the main focus of this paper.

In the following sections, we perform systematic numerical analyses to quantify any increase in the efficiency of parameter estimation by using Bayesian fusion of EEG and fMRI relative to unimodal inversion of fMRI data alone. In brief, we first simulate synthetic EEG and fMRI timeseries under the same task and neuronal model. We then implement two inversion schemes for synthetic multimodal data, corresponding to Bayesian fusion (following the pipeline above) and inversion using only fMRI data. Finally, we introduce information gain to quantify the relative utility of each modality and the data assimilation enabled by Bayesian fusion.

*Simulations: multimodal data for mismatch negativity paradigm*

We simulated EEG and fMRI timeseries for a widely used mismatch negativity (or auditory oddball) paradigm that has been extensively explored using DCM (Garrido et al., 2007; Kiebel et al., 2007a; Garrido et al., 2009a; Garrido et al., 2009b; Kiebel et al., 2009; Auksztulewicz and Friston, 2015). To ensure the synthetic data were biologically plausible, we first conducted a standard DCM analysis using empirical, open access data (https://www.fil.ion.ucl.ac.uk/spm/data/eeg_mmn/). We used the ensuing estimates of effective connectivity and synaptic (and observation noise) parameters to generate synthetic data.

Early DCM studies of the mismatch negativity identified a hierarchical network of five cortical sources to explain evoked responses elicited by standard and oddball stimuli. These sources included left and right primary auditory cortex (left A1, right A1), left and right superior temporal gyrus (left

STG, right STG), and right inferior frontal gyrus (right IFG) (Garrido et al., 2007; Kiebel et al., 2007a; Garrido et al., 2009a; Kiebel et al., 2009). In our simulations, we defined a hierarchical architecture comprising these five sources: the sources were hierarchically connected via extrinsic forward and backward connectivity. The extrinsic forward connections arose from superficial pyramidal cells of left A1, right A1 and right STG, and targeted spiny stellate cells and deep pyramidal cells of left STG, right STG and right IFG, respectively. The extrinsic backward connections arose from deep pyramidal cells of right IFG, right STG and left STG, to superficial pyramidal cells and inhibitory interneurons of right STG, right A1 and left A1, respectively (see the dark blue connections in Fig. 2B).

In addition to these extrinsic connections, there are intrinsic inhibitory, and excitatory connections within and among the four populations that constitute the canonical microcircuit for each source (the pink and light blue connections in Fig. 2B). External experimental stimuli (i.e., driving inputs) arrived at the spiny stellate populations in bilateral auditory sources. Modulatory effects (i.e., modulatory inputs), which were responsible for explaining the differences in responses to standard and oddball conditions, were specified by changes in: (i) extrinsic forward connectivity from superficial pyramidal cells of left A1, right A1 and right STG, to spiny stellate cells of left STG, right STG and right IFG, respectively, (ii) extrinsic backward connectivity from deep pyramidal cells of right IFG, right STG and left STG, to superficial pyramidal cells of right STG, right A1 and left A1, respectively, and (iii) self-inhibitions of superficial pyramidal cells in bilateral auditory sources (see Fig. 2B). The precise details of this model of the mismatch negativity are not crucial for the purposes of this paper; however, we ensured that the model generating these synthetic data was as close as possible to real neuronal networks – and their context or condition sensitive connectivity.

After specifying this model, we conducted a DCM analysis using the empirical EEG data, and recovered a total of 74 biologically plausible parameter estimates, which included 12 extrinsic connectivity parameters in extrinsic connectivity $\mathbf{A}$ matrices, 8 modulating effect parameters in $\mathbf{B}$ matrices, 2 driving effect parameters in vector $\mathbf{C}$, an intrinsic gain parameter in $\mathbf{S}$, 20 self-inhibition parameters in $\mathbf{G}$, 20 synaptic rate constant parameters in $\mathbf{T}$, and 11 conduction delay parameters in $\mathbf{D}$. The posterior expectations of these empirically determined neuronal parameters were then used as the parameter values generating the neuronal dynamics in the subsequent generation of synthetic EEG and fMRI data. The connectivity architecture and the true parameter values are shown in Fig. 2B.

For the generation of synthetic EEG, in addition to the neuronal parameters, we specified EEG spatial parameters using the lead field employed in the above DCM analysis of empirical data. These included three location and three orientation parameters for each source (the values of these spatial parameters are shown in Fig. 2B). The experimental (auditory) driving input was modelled as a Gaussian shape function of peristimulus time (see Fig. 2A). Finally, we used the forward model of the canonical microcircuit DCM for EEG to generate sensor data for standard and oddball conditions, respectively. The simulated sensor signals with observation noise served as the synthetic EEG data in subsequent Bayesian fusion analyses. A plausible level of observation noise was created by convolving Gaussian noise (with a standard deviation of one half of the generated response amplitude) with a smoothing kernel of eight time bins (4 ms). The resulting synthetic EEG data are shown in Fig. 3A.

Synthetic fMRI data were generated using the same neuronal parameters by specifying the timing of events in a simulated event-related fMRI of the auditory oddball paradigm, as well as a set of appropriate haemodynamic parameters. To ensure the synthetic multimodal data corresponded to the same experimental paradigm (and the two data modalities were generated by the same neuronal responses), we used the EEG mismatch negativity paradigm to derive an event-related auditory oddball paradigm for fMRI. More precisely, we read the trial definition file of the empirical EEG data to recover the stimulus onset times (SOTs) of all events (80% of standard events and 20% of oddball events in a pseudo-random sequence, with inter-stimulus interval of 1.5 s, see Fig. 2A). We then added a random jitter (of -0.25 to 0.25 s, uniformly distributed) to define the SOTs for the event-related fMRI paradigm (Kiebel et al., 2007b).Given that the neuronal responses are driven by all auditory stimuli, the modulating effects were restricted to oddball events. We modelled the driving input as a stick function encoding the SOTs of all events, and encoded the modulatory input as a box-car function (a box-car function allowed the changes in connectivity to persist for the duration of neuronal transients) following the SOTs of oddball events (see Fig. 2A). The TR was set to 1.7 s to emulate a typical repetition time. The haemodynamic parameters were sampled from their prior densities in Table 2, which included 12 neurovascular coupling parameters denoted by **J** and 11 haemodynamic parameters denoted by **H** (the values are shown in Fig. 2B) for the five regions. After specifying the experimental inputs, the neuronal parameters and the haemodynamic parameters, we generated BOLD responses in the five regions. We then added appropriate level of observation noise (a signal-to-noise ratio of four, noise was generated using the SPM Matlab code **spm_dcm_generate**.m ) to the simulated BOLD responses (Kiebel et al., 2007b), and finally took a section of resulting timeseries – with length of 256 scans – from the noisy BOLD responses to be the synthetic fMRI data used in subsequent analyses. The ensuing synthetic fMRI timeseries are shown in Fig. 3B.

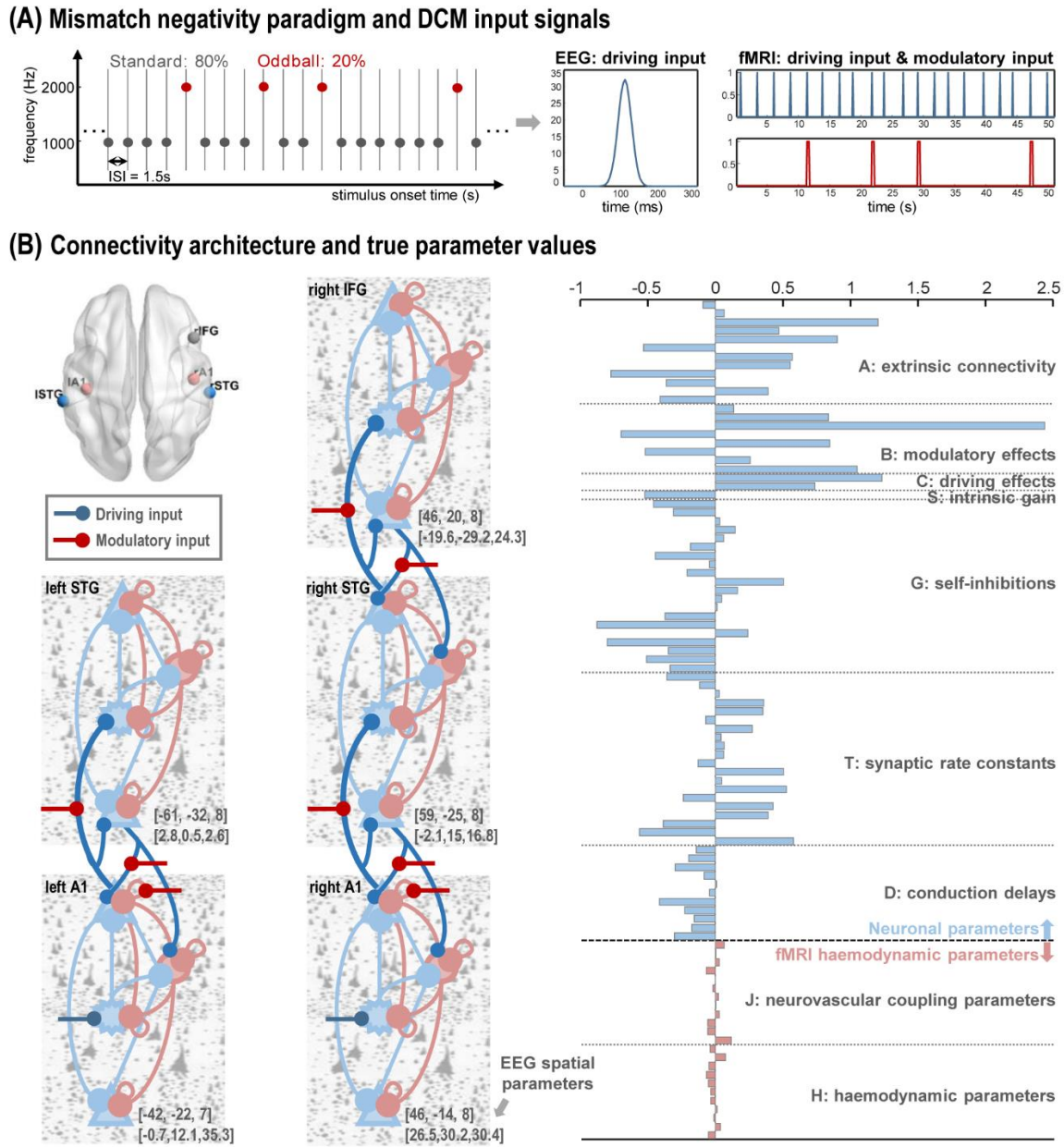

**Fig. 2.** Experimental inputs, connectivity architecture and true parameter values used for simulations. (A) Mismatch negativity paradigm and DCM input signals. The mismatch negativity paradigm contains standard (1000 Hz) and oddball tones (2000 Hz), occurring 80% and 20% respectively in a pseudo-random sequence, with an inter-stimulus interval (ISI) of 1.5 s. The DCM driving input for EEG data is modelled as a Gaussian bump function of peristimulus time, while – for fMRI – the driving input is a stick function encoding stimulus onset times (SOTs) of all tones, where the modulatory input is a box-car function following the SOTs of oddball events. (B) Connectivity architecture and true parameter values. This architecture comprises five regions; including left and right A1 (A1 = primary auditory cortex), left and right STG (STG = superior temporal gyrus) and right IFG (IFG = inferior frontal gyrus), visualized using the BrainNet Viewer (Xia et al., 2013). Each region is a canonical microcircuit with inhibitory (pink) and excitatory (light blue) intrinsic connections, where the five regions are coupled via extrinsic forward and backward connections (dark blue). The driving inputs (cyan) target spiny stellate cells in bilateral A1 regions, and the modulatory inputs (red) are specified by changes in either intrinsic or extrinsic connections. The true values of neuronal and haemodynamic parameters used for generating synthetic multimodal data are shown using a bar chart.

*Model inversion schemes: Bayesian fusion vs. fMRI only inversion*

In this section, we focus on the comparison of fMRI inversion with and without Bayesian fusion. We first performed Bayesian fusion using both synthetic EEG and fMRI data. Our procedure was as follows:

1. Specification of priors for EEG: we specified the connectivity model (the connectivity architecture shown in Fig. 2B), and the uninformative priors over neuronal parameters according to Table 1. In addition, we placed precise priors over the EEG spatial (lead field) parameters, based upon the spatial parameters (see Fig. 2B) used to generate the synthetic EEG data. This ensured the neuronal posteriors from the EEG analysis were not confounded by conditional dependencies between EEG spatial and neuronal parameter estimates. (See Kiebel et al. (2006) for an evaluation of the effects of lead field specification on the recovery of model parameter estimates).

2. Model inversion for EEG: we performed a DCM analysis for EEG using the above priors, to obtain the posterior means and covariances for neuronal parameters (inverted using the SPM Matlab code **spm_nlsi_N**.m).

3. Specification of priors for fMRI: we specified the connectivity model (connectivity architecture shown in Fig. 2B), and the uninformative priors over haemodynamic parameters (Table 2), but replaced the prior means and covariances of the neuronal parameters with the posterior means and covariances from the EEG inversion (i.e., **Bayesian belief updating**).

4. Model inversion for fMRI: we conducted a DCM analysis for fMRI using the above priors, to obtain the posterior beliefs over all free parameters informed by both EEG and fMRI data (inverted using **spm_dcm_fmri_nmm**.m).

We also performed fMRI only inversion for comparison purposes. The only difference between the two fMRI inversions concerned the priors over neuronal parameters: the fMRI only inversion retained the uninformative neuronal priors (Table 1), as opposed to the empirical neuronal priors derived from the EEG analysis. Using different priors causes the Variational Laplace scheme to start from different initialisation points (i.e., from prior expectations) during parameter estimation. To ensure the results of the two inversion schemes could not be explained by different initialisations, we used the starting points of Bayesian fusion to initialise the fMRI only inversion. In summary, the fMRI only inversion comprised the following steps:

1. Specification of priors for fMRI: we specified the connectivity model (the connectivity architecture shown in Fig. 2B), and the uninformative priors over neuronal and haemodynamic parameters according to Table 1 and Table 2, respectively.
2. Model inversion for fMRI: we conducted a DCM analysis for fMRI using the above priors and the initialisation point of Bayesian fusion (by setting **DCM.options.P**), to obtain the posterior beliefs over all free parameters (inverted using **spm_dcm_fmri_nmm**.m).

After implementing the two inversion schemes, we evaluated the quality of model inversions and the efficiency of parameter estimation from the following three perspectives: (i) we conducted Bayesian model comparison based on the (negative) free energy (as a bound approximation to log model evidence), (ii) checked the consistency between the predicted and true (synthetic) BOLD responses, (iii) compared the reduction in posterior variances of the parameter estimates, which scores the shrinkage of uncertainty. The free energy and the parameter estimation results are shown in Fig. 5 and Fig. 6.

*Quantification of information gain*

To further quantify the relative utility of EEG and fMRI in making inferences about various model parameters – and to evaluate any increase in the efficiency of parameter estimation attributable to Bayesian fusion – we calculated the information gain. This measure of data quality is the Kullback-Leibler (KL) divergence between prior and posterior beliefs, after observing one or more data modalities. Under the Laplace assumption employed in DCM; i.e., the prior and posterior probabilities are Gaussian, we used the KL divergence between multivariate Gaussians:

$$D(N_1 \parallel N_0) = \frac{1}{2}\left(\text{tr}(\Sigma_0^{-1}\Sigma_1) + (\mu_0 - \mu_1)^{\text{T}}\Sigma_0^{-1}(\mu_0 - \mu_1) - k + \ln\frac{\det \Sigma_0}{\det \Sigma_1}\right)$$

Where $N_0 = \text{N}(\mu_0, \Sigma_0)$ denotes the prior density over parameters with mean $\mu_0$ and covariance $\Sigma_0$, $N_1 = \text{N}(\mu_1, \Sigma_1)$ denotes the corresponding posterior density, and $k = \text{rank}(\Sigma_0)$ indicates the number of free parameters. The divergence $D(N_1 \parallel N_0)$ measures the difference (in units of nats) from the prior density $N_0$ to posterior probability density $N_1$, which in the context of Bayesian inference reflects the information gained when we revise our beliefs about parameters (by observing data). In other words, it scores the degree of belief updating afforded by any given data (Kullback and Leibler, 1951; Burnham and Anderson, 2003; Duchi, 2007; Zeidman et al., 2018).

Therefore, based on the original prior ($p(\theta|m)$), the unimodal posteriors for EEG ($p(\theta|Y_{EEG})$) and fMRI ($p(\theta|Y_{MRI})$), as well as the posterior given by Bayesian fusion ($p(\theta|Y_{MRI}, Y_{EEG})$), we evaluated the following information gain:

1. $D[p(\theta|Y_{MRI}) \parallel p(\theta|m)]$ – The KL divergence from original priors to posteriors given only fMRI data, to assess to what extent the fMRI data could inform parameter estimation. For convenience, we denoted this KL divergence as **D1**. This KL divergence can be decomposed with respect to the neuronal and haemodynamic parameters separately: we accordingly report the results as **D1$_N$** for neuronal part and **D1$_H$** for haemodynamic part, respectively.
2. $D[p(\theta|Y_{EEG}) \parallel p(\theta|m)]$ – The KL divergence from original priors to posteriors given only EEG data, to quantify the amount of information that EEG data contributes in Bayesian fusion analysis. We report this KL divergence as **D2**, **D2$_N$** and **D2$_H$**.
3. $D[p(\theta|Y_{MRI}, Y_{EEG}) \parallel p(\theta|Y_{EEG})]$ – The KL divergence from posteriors given only EEG data to posteriors given both modalities, to quantify the amount of information that the subsequent fMRI data could provide after observing EEG data (i.e., Bayesian fusion). We denoted this KL divergence as **D3**, **D3$_N$** and **D3$_H$**.

We used these information gain results to quantify the amount of information contained in fMRI, EEG and their fusion. Based on our previous assumption that EEG would be more informative than fMRI – with respect to neuronal parameters – we would expect **D2$_N$** to be greater than **D1$_N$**. If Bayesian fusion accounts for conditional dependencies between neuronal and haemodynamic parameters, a better estimate of haemodynamic parameters would be obtained – as reflected here by an expected greater value of **D3$_H$** than **D1$_H$**. We report the information gain using bar charts, as shown in Fig. 7.

## Results

### *The simulated multimodal data*

The simulated multimodal data are shown in Fig. 3, which reproduce plausible responses to the mismatch negativity paradigm in EEG research and the auditory oddball design in fMRI. The upper panel of Fig. 3A shows the synthetic EEG sensor data (127 channels, with observation noise) for the standard and oddball conditions respectively. The lower panel shows the first principal component of the sensor data for the two conditions and their difference, as well as the observation noise added to the sensor responses. The accompanying simulated BOLD response (with observation noise) of each of the five regions is respectively shown in Fig. 3B. These synthetic fMRI timeseries reflect a succession of frequent standard stimuli and occasional oddball stimuli and show that different regions exhibit functionally selective responses to oddball events.

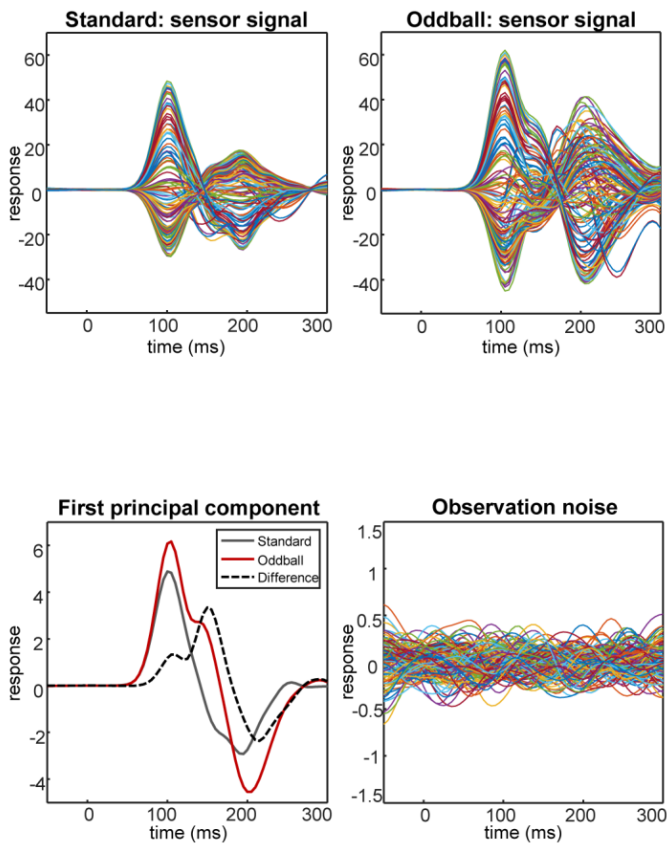
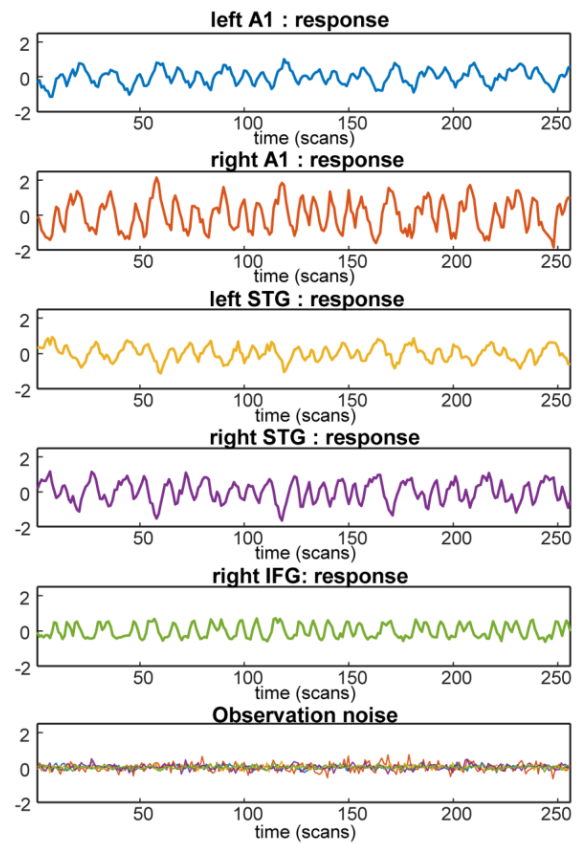

**Fig. 3.** The simulated multimodal data. (A) The synthetic EEG data. The upper panel shows the synthetic EEG sensor data (127 channels, with observation noise) for the standard and oddball conditions respectively. The lower panel shows the first principal components of the sensor data for the two conditions and their difference (i.e., mismatch negativity), and the observation noise added to the sensor data. (B) The synthetic fMRI data. The simulated BOLD response (with observation noise) of each of the five regions is shown in this panel.

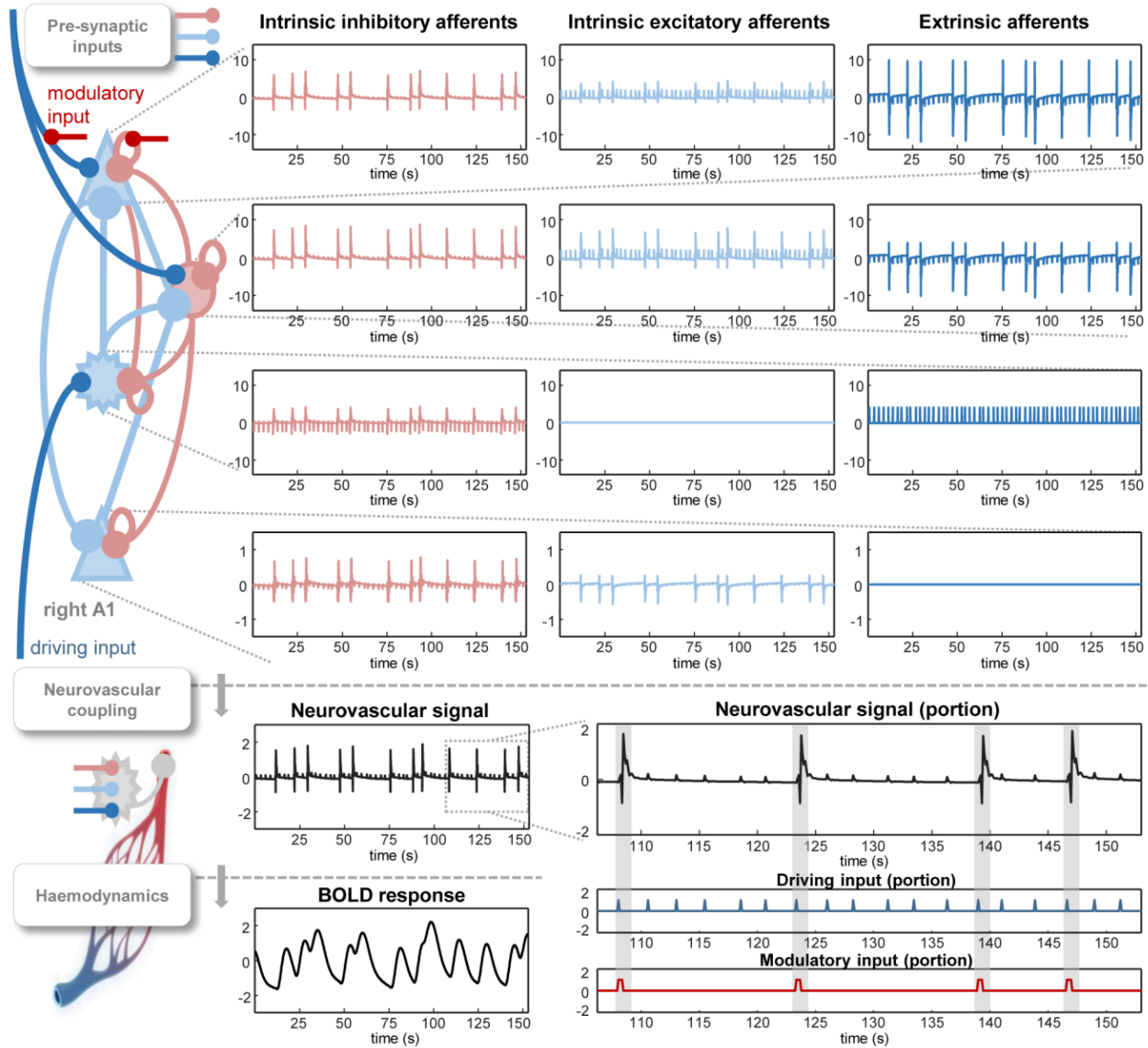

**Fig. 4.** The predicted laminar-specific responses and neurovascular signal (of one brain region: right A1) when modelling fMRI with the canonical microcircuit DCM. Each of the four populations receives three sorts of pre-synaptic inputs, corresponding to intrinsic inhibitory afferents (pink), intrinsic excitatory afferents (light blue), and extrinsic afferents (dark blue). The subsequent neurovascular signal is the linear mixture (scaled by the neurovascular coupling parameters) of these pre-synaptic inputs. This signal is then magnified to show in detail how experimental effects (i.e., driving and modulatory inputs) are expressed at the synaptic level. The final BOLD signal is generated by the haemodynamic convolution of the neurovascular signal, where the BOLD response fluctuations correspond to oddball events.

To provide insight into the laminar specific responses measured with fMRI, we show an example in Fig. 4 of the simulated pre-synaptic inputs and the neurovascular signal of one brain region (right A1). The upper panel of Fig. 4 shows the three sorts of pre-synaptic inputs (i.e., intrinsic inhibitory afferents, intrinsic excitatory afferents and extrinsic afferents) to each of the four populations, where the 12 pre-synaptic inputs exhibit different amplitudes (some of them are zero indicating no such afferent) but similar patterns. The consequent neurovascular signal – shown in the lower panel – is a linear mixture (scaled by the 12 neurovascular coupling parameters) of these pre-synaptic inputs. The signal is then magnified to show how experimental effects (i.e., driving and modulatory inputs) are expressed at the synaptic level: here, we see that every event onset triggers a laminar response, while the oddball events cause much stronger fluctuations in comparison with standard events. The final BOLD signal is generated by a haemodynamic convolution of the neurovascular signal. The resulting BOLD responses correspond to oddball events, because responses to standard events are smoothed over time by the haemodynamics. This also suggests that the neuronal parameters that can be estimated relatively more efficiently, using fMRI, are the changes in connectivity that generate oddball responses.

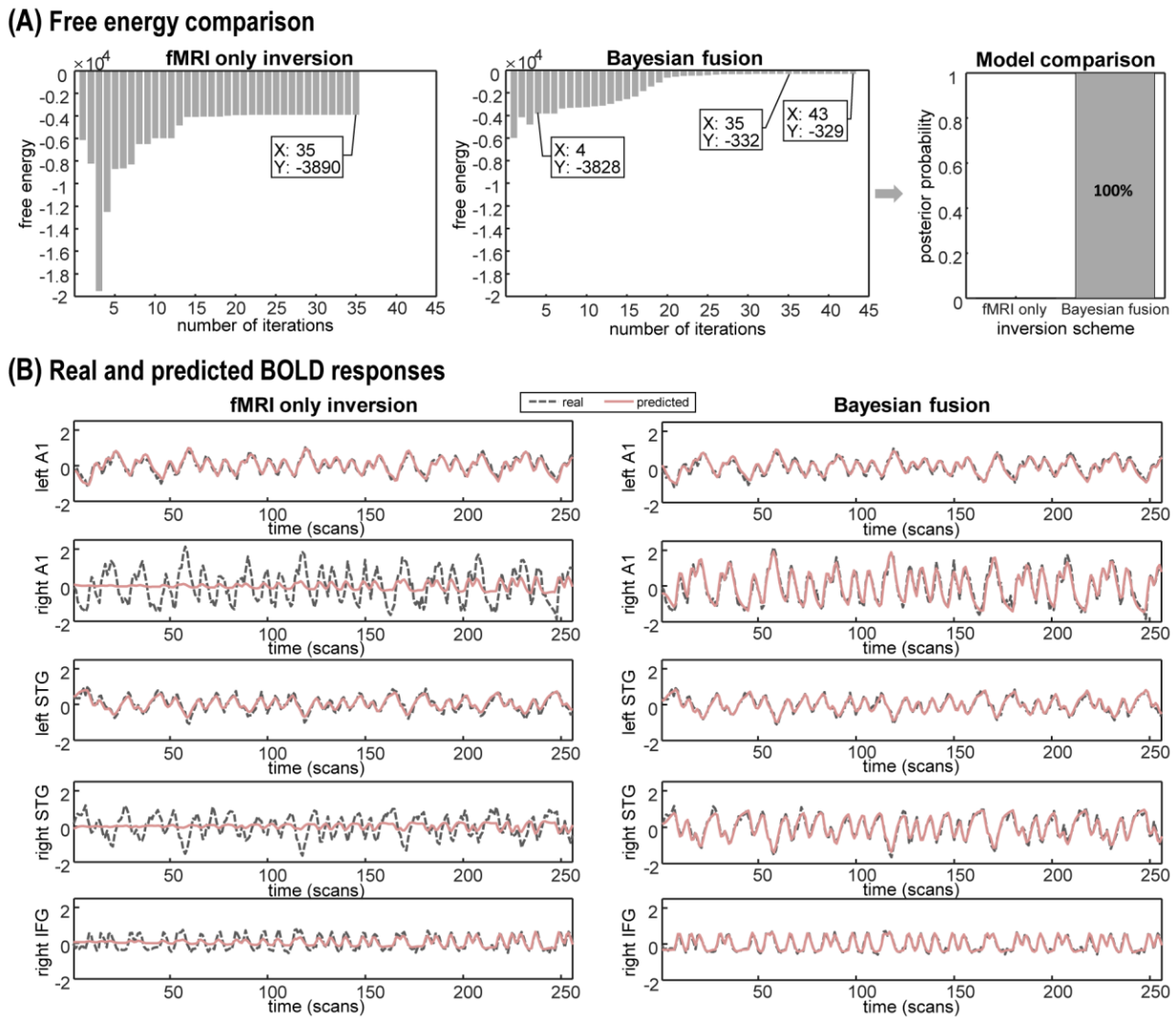

**Fig. 5.** Comparison of model inversion results between fMRI only inversion and Bayesian fusion. (A) The free energy changes over iterations of a Variational Laplace scheme, and the ensuing Bayesian model comparison of the two inversion schemes. (B) The real (grey dotted lines) and predicted (pink solid lines) BOLD responses correspond to the two inversion schemes.

*Model inversion results*

Fig. 5 compares the free energy and the consistence between real and predicted BOLD responses under the two inversion schemes; namely, the fMRI only inversion and Bayesian fusion. In Fig. 5A one can see that, in general, the free energy (i.e., log model evidence) increases with increasing number of iterations in both inversion schemes, endorsing the Variational Laplace scheme used to estimate model parameters and evidence. In the fMRI only inversion, the scheme converges after 35 iterations (to a free energy of -3890 nats), while in Bayesian fusion, the free energy reaches a plateau after 27 iterations, and converges at the 43$^{rd}$ iteration (to a much larger free energy of -329 nats). The subsequent Bayesian model comparison indicates, as expected, that Bayesian fusion is better than the fMRI only inversion with posterior probability of 100%. Fig. 5B shows the real and predicted BOLD responses under the two inversion schemes; in which the real BOLD signals correspond to the grey dotted lines, and predicted BOLD signals are denoted by pink solid lines. These results show that, with Bayesian fusion, the predicted BOLD signals of all the five regions match the real signals remarkably well, but for the fMRI only inversion, we see obvious inconsistencies between the predicted and real BOLD in the right A1, right STG and right IFG, indicating a relatively poor parameter estimates.

Fig. 6 shows the prior and posterior beliefs about the parameters (four both neuronal and haemodynamic parameters) as well as the conditional correlations among these parameters; for the fMRI only inversion (Fig. 6A) and Bayesian fusion (Fig. 6B). We first compare the prior beliefs and correlation matrices of the two inversion schemes. For fMRI only inversion, the default priors for both neuronal and haemodynamic parameters are uninformative (zero means and large variances) and independent of each other (i.e., a diagonal form for the prior correlation matrix). After Bayesian belief updating, the EEG modality moves the neuronal priors away from a mean of zero, and reduces the prior variance of most neuronal parameters; these reduced values reflect a reduction in uncertainty

**(A) fMRI only inversion**

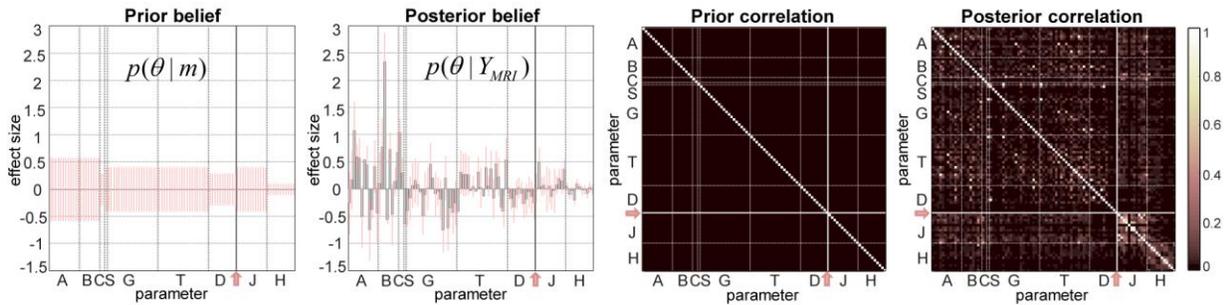

**(B) Bayesian fusion**

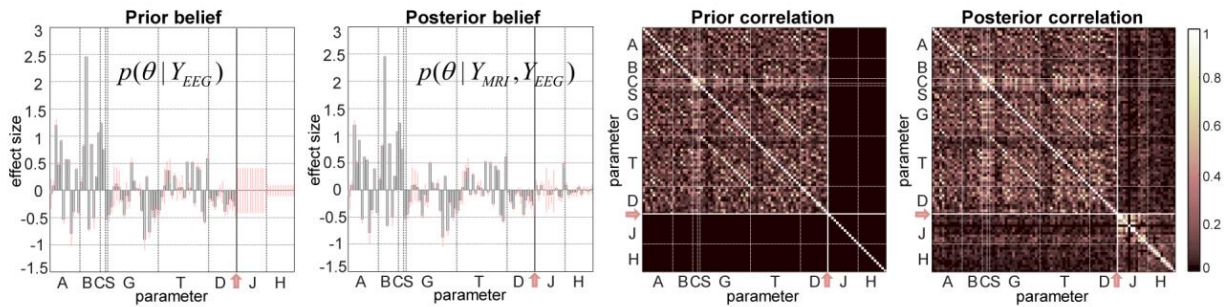

**Fig. 6.** The prior and posterior beliefs of the model parameters (see Fig. 2, Table 1 and Table 2 for a detailed description of each model parameter), as well as the conditional correlations among these parameters, corresponding to (A) fMRI only inversion and (B) Bayesian fusion of EEG and fMRI.

about neuronal parameters after observing EEG data. In addition, the (empirical) prior correlation matrix over the neuronal parameters is no longer diagonal, reflecting conditional dependencies given informative electromagnetic data.

The subsequent posterior beliefs and posterior correlation matrices for fMRI only inversion and Bayesian fusion, respectively, reveal the informativeness of fMRI data in estimating neuronal and haemodynamic parameters before and after seeing the EEG modality. Here, we were particularly interested in the efficiency with which haemodynamic parameters could be estimated after observing EEG data. Clearly, we obtain no direct information about haemodynamic parameters following inversion of EEG data: this is reflected in the fact that the prior beliefs over haemodynamic parameters are the same under the two inversion schemes. However, it is possible that EEG data can provide additional information about haemodynamic parameters, in virtue of conditional dependencies. In other words, EEG data can resolve uncertainty about neuronal parameters – after seeing fMRI data – if, and only if, neuronal and haemodynamic parameter estimates depend on each other. Indeed, by inspecting the posterior correlation matrices in both Fig. 6B and Fig. 6A, one can see clear conditional correlations between neuronal and haemodynamic parameters. This means that reducing uncertainty about neuronal parameters will affect uncertainty about haemodynamic parameters – allowing for a synergy between EEG and fMRI data during model inversion. We now quantify this kind of synergetic interaction using the following information gain results.

*Information gain*

Fig. 7 shows the information gain results as described above. The upper panel shows the amount of information we have obtained by conducting the fMRI only inversion, while the lower panel illustrates the information accumulation from default priors to EEG posteriors (i.e., Bayesian belief updating), then from EEG posteriors to bimodal posteriors in the Bayesian fusion of EEG and fMRI. Overall, Bayesian fusion reveals more information than just inverting fMRI data (i.e., $\mathbf{D2} + \mathbf{D3} > \mathbf{D1}$ )**,** and this increase in information gain is reflected in both neuronal and haemodynamic decompositions. More specifically, in terms of the neuronal parameters, we obtain greater value of $\mathbf{D2_N}$ than $\mathbf{D1_N}$ suggesting that EEG is more informative than fMRI. This endorses our understanding of the results in Fig. 6; namely, that EEG data contains much more information about neuronal parameters. For the haemodynamic parameters, we obtain no information following inversion of EEG data (i.e., $\mathbf{D2_H} = \mathbf{0}$). However, using the empirical priors from EEG, the fMRI data provides more information about the parameters that mediate neurovascular coupling, relative to just observing fMRI (i.e., $\mathbf{D3_H} > \mathbf{D1_H}$). This increase in information gain furnishes quantitative evidence that Bayesian fusion of EEG and fMRI can resolve conditional dependencies between neuronal and haemodynamic parameters; thereby improving the efficiency of parameter estimation based on non-invasive fMRI data and neural mass models of the canonical microcircuit.

## Discussion

In this paper, we evaluated the contribution of multimodal EEG and fMRI data for estimating neuronal architectures using the canonical microcircuit DCM. In summary: 1) we created biologically plausible (based on empirically determined parameters) simulations of multimodal neuroimaging data (i.e., EEG and BOLD responses, see Fig. 3) under the same paradigm (i.e., the mismatch negativity paradigm, see Fig. 2). Crucially, these multimodal data were generated using a common neuronal architecture (Fig. 1); 2) we simulated laminar-specific responses (i.e., pre-synaptic inputs) and ensuing neurovascular signal, which encoded how experimental effects (i.e., driving and modulatory inputs)

**(A) fMRI only inversion**

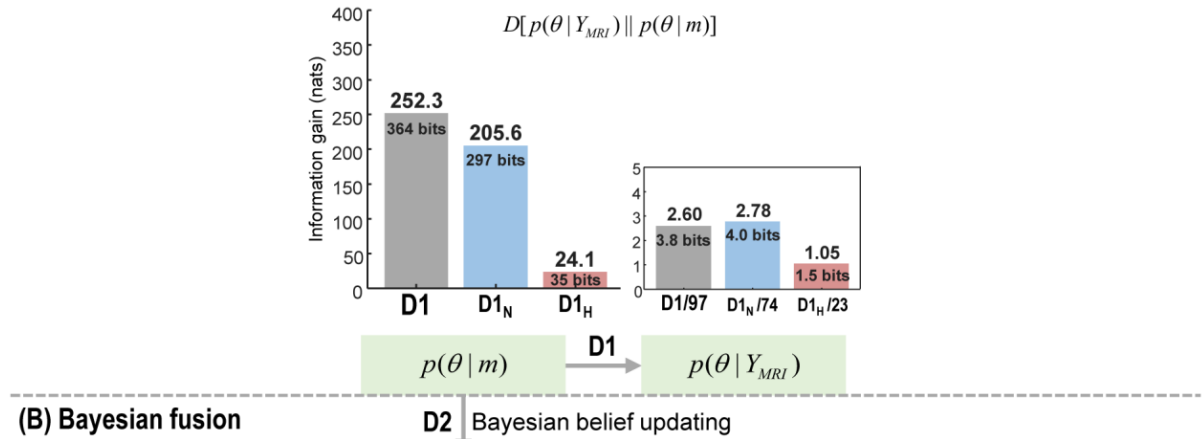

**(B) Bayesian fusion**

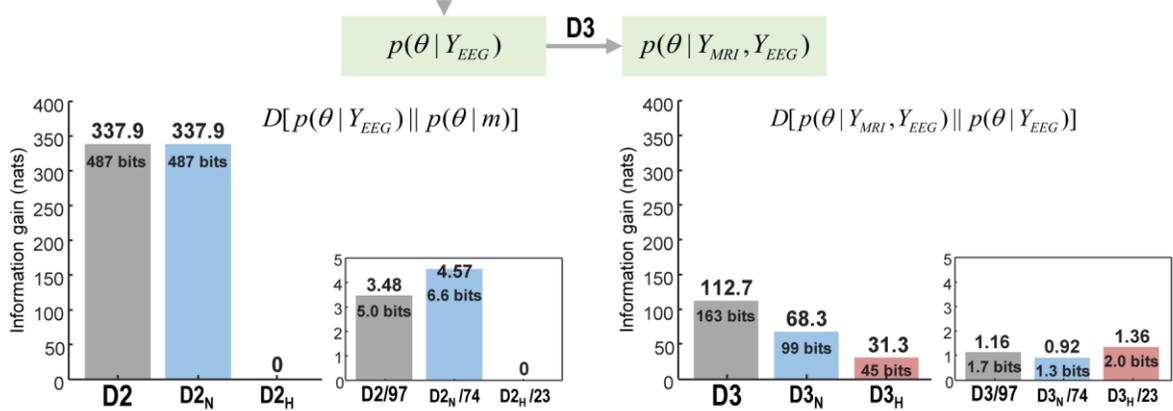

**Fig. 7.** Information gain (in units of nats or bits, 1 nat = $\mathbf{log_2(e)}$ bits) associated with fMRI and EEG data separately – and their fusion. (A) The KL divergences from default priors to posteriors, given only fMRI data (i.e., D1), and the corresponding neuronal and haemodynamic decompositions (i.e., D1$_N$ and D1$_H$ respectively). This set of KL divergences speak to the information gain obtained in fMRI only inversion. (B) The KL divergences from default priors to empirical priors, given only EEG data (i.e., D2, D2$_N$ and D2$_H$), and then from empirical priors given EEG to posteriors, given both EEG and fMRI (i.e., D3, D3$_N$ and D3$_H$). This set of KL divergences quantify the information gain under multimodal Bayesian fusion. These KL divergences are divided by the number of parameters (i.e., 97 model parameters, including 74 neuronal parameters and 23 haemodynamic parameters), to measure the information gain afforded by each parameter.

are expressed at the synaptic level (see Fig. 4); 3) comparative analyses showed that Bayesian fusion furnished better model parameter estimation, which was reflected by increases in free energy (i.e., log model evidence), a better match between real and predicted BOLD signals (see Fig. 5), and a reduction of posterior variance, indicating a shrinkage of uncertainty about model parameters (see Fig. 6); 4) and information gain provided quantitative evidence that Bayesian fusion can leverage conditional dependencies between neuronal and haemodynamic parameters; thereby evincing a synergetic resolution of uncertainty about model parameters (see Fig. 7).

This paper is the first formal demonstration that Bayesian fusion of EEG and fMRI supports inferences about detailed changes in microcircuitry – and characterisations of laminar specific neurovascular coupling, as measured with non-invasive fMRI data. These results also strengthen our understanding of the canonical microcircuit DCM for fMRI (Friston et al., 2017). Based on this modelling framework and Bayesian fusion, one could propose and test many hypotheses pertaining to laminar specific cortical architectures, such as exploring microcircuit models of specific regions or hierarchical networks, with distinct forward and backward extrinsic connections. One could also evaluate experimental effects (e.g., attention, visual perception, pharmacologic, etc.) on changes in either extrinsic (long-range) or intrinsic (short-range) connections at the level of specific neuronal populations: e.g., superficial or deep pyramidal cells (Heinzle et al., 2007; Brown and Friston, 2012; Auksztulewicz and Friston, 2015; Tsvetanov et al., 2016; Friston et al., 2017; Havlicek et al., 2017; Lawrence et al., 2018). These sorts of questions shift our focus away from simply localising regional responses (using fMRI) towards an ever more detailed characterisation of functional integration and physiologically grounded research that exploits the increasing temporal (M/EEG) and spatial (fMRI) fidelity of neuroimaging data.

This paper has focused on the technical procedures entailed by this kind of Bayesian fusion; however, potential domains of application deserve some comment. In one sense, people already use Bayesian fusion when they use (empirical) priors from the location of fMRI activations for source reconstruction in EEG. This is particularly the case in dynamic causal modelling of EEG data where the prior location of equivalent current dipoles (ECDs), modelling each electromagnetic source, requires location (i.e., spatial) priors – that are often based on fMRI (Daunizeau et al., 2007; Henson et al., 2010). The Bayesian fusion described in this paper completes the synergetic use of both modalities by enabling the posterior estimates from EEG – that inherit spatial information from fMRI – to inform the modelling of fMRI – so that it inherits temporal information from EEG. One of the key benefits of this is that one can resolve conditional dependencies between neuronal and haemodynamic parameters. This is an important observation because it speaks to key questions about neurovascular coupling and attributing differences in evoked or induced responses to changes in neuronal connectivity or haemodynamics. For example, is the effect of ageing on fMRI responses attributable to changes in neuronal architectures, changes in the elasticity of blood vessels – or both? e.g., Fabiani et al. (2014). In principle, the scheme described in this paper should provide an optimal estimate of changes in neuronal and haemodynamic parameters to address this sort of question.

The multimodal Bayesian fusion proposed in this paper – particularly in reference to using posterior estimates from DCM analysis of EEG as empirical priors for subsequent DCM analysis of fMRI – rests upon a common neuronal model; namely, the canonical microcircuit neural mass model. This neuronal model was developed for DCM for M/EEG then applied to DCM for fMRI in (Friston et al., 2017). However, the notion of adopting the same neuronal model to explain both data modalities was discussed in the foundational paper describing DCM for M/EEG (David et al., 2006), in which the authors pointed out that the most significant challenge of employing over-parameterised neural mass models for fMRI data was guaranteeing an efficient model inversion. In DCM, model inversion generally uses standard Variational Laplace procedures to estimate model parameters and evidence for inferences about specific connections and network structure; where the model evidence corresponds to accuracy minus complexity (Penny et al., 2004; Friston et al., 2007). In other words, a model with too many parameters is a model whose complexity cost exceeds the increase in accuracy or goodness of fit afforded by extra parameters. In this study, we used two model inversion schemes; namely, the Bayesian fusion of EEG and fMRI, and the fMRI only inversion, and demonstrate clearly that the fMRI only inversion is inefficient in relation to Bayesian fusion. Some 'flat line' predictions of BOLD responses in fMRI only inversion, as well as the convergence of free energy to a much lower value compared with that in Bayesian fusion (see Fig. 5), suggest that the parameter estimation of canonical microcircuit DCM is inefficient given merely fMRI data and without the empirical priors afforded by EEG. Interestingly, the complexity part of (log) evidence is exactly the information gain used to assess the contribution from different data modalities above. Heuristically, informative data 'pulls' the posterior density – over unknown parameters – away from the prior density to provide an accurate account of the data (i.e., maximise model fit). In other words, the accuracy 'pays for' a complexity cost, which is the information gain. Model evidence is therefore the difference between the accuracy and complexity or information gain.

The information gain afforded by different modalities have a useful quantitative interpretation. They are measured in natural units (nats) that can be converted into bits by multiplying with $\log_2(e)$. One bit of information allows us to say whether a particular parameter is high or low. For example, the information gain about the neuronal parameters, given the EEG data (i.e., **D2$_N$**) is about 487 bits. If we distribute this information over the 74 neuronal parameters, this suggests that we gain about six bits of information about each parameter. In other words, we could place the strength of each connection or synaptic rate constant in one of six 'bins'. The information gain – afforded by Bayesian fusion – about the haemodynamic parameters (i.e., **D3$_H$**) is about 31 nats or 45 bits. Given we used 23 haemodynamic parameters; this corresponds to about two bits per parameter (see Fig. 7).

Some readers may wonder why inverting the canonical microcircuit DCM using just fMRI (without any information from EEG) was possible in Friston et al. (2017). In that paper, the authors fixed the neuronal parameters (including the synaptic rate constants, self-inhibitions, and conduction delays) to their prior mean, when conducting model inversion; thereby reducing complexity. This contrasts with our fMRI only inversion, where we inverted the synthetic fMRI data without fixing any neuronal parameters – by doing this we leveraged the information gain afforded by Bayesian belief updating. In Friston et al. (2017), the neuronal parameters were fixed to deflate the model complexity to ensure more efficient balance between accuracy and complexity. This illustrates the fact that priors can have an important role in determining the efficiency of model inversion (Kiebel et al., 2006); especially in the context of canonical microcircuit DCM for fMRI. In practice, canonical microcircuit DCM analyses with fMRI therefore require a careful consideration of model complexity – and the provision of precise constraints on model parameters; such as the empirical priors afforded by Bayesian fusion.

Several lines of research suggest themselves for future work. First, it will be useful to establish which applications would benefit from inverting a single generative model of EEG and fMRI data, as presented here, and which applications could be addressed by fitting separate neuronal and haemodynamic models to each modality, for example, using neuronal responses from a DCM analysis of EEG to drive a haemodynamic model of fMRI responses (Jafarian et al., 2019). In principle, this question can be addressed using Bayesian model comparison. Second, the utility of the canonical microcircuit DCM and multimodal Bayesian fusion as a clinical computational assay or phenotyping to characterise pharmacological, pathophysiological and cognitive dysfunctions using patient data (Boly et al., 2011; Campo et al., 2012; Uludag and Roebroeck, 2014; Benetti et al., 2015; Mechelli et al., 2017). Additionally, when dealing with multi-subject and multi-model Bayesian fusion, the parametric empirical Bayes (PEB) procedures may be useful (Henson et al., 2011; Friston et al., 2015; Litvak et al., 2015; Wakeman and Henson, 2015; Friston et al., 2016). Third, the refinement of the current generative model, especially optimizing the parameterisation of neurovascular coupling, may benefit from carefully selected empirical data; while using Bayesian fusion to resolve conditional dependencies between neuronal and haemodynamic parameters (Friston et al., 2017). Finally, the expansion of the current task-based analysis to the corresponding resting-state methodology (Friston et al., 2014; Razi et al., 2017), where an equivalent canonical microcircuit formulation for cross spectral data features, e.g., Razi et al. (2015), will be needed.

## Acknowledgements

DWH is funded by the National Natural Science Foundation of China (61420106001) and the National Key Research and Development Program of China (2018YFB1305101). KJF is funded by the Wellcome Trust Principal Research Fellowship (Ref: 088130/Z/09/Z). HLW is funded by the China Scholarship Council (201703170238).

**Declarations of interest**

None.